\documentclass[prodmode,acmtosn]{acmsmall}

\usepackage{cite}
\usepackage{graphicx}
\usepackage{url}
\usepackage{amsmath,amssymb,amsfonts}
\usepackage{algorithmicx}
\usepackage{algorithm}
\usepackage{algpseudocode}
\usepackage{subfigure}

\newcommand\etal{\emph{et al.}}

\title{In-Network Distributed Solar Current Prediction}

\author{ELIZABETH BASHA
\affil{Massachusetts Institute of Technology, University of the Pacific}
RAJA JURDAK
\affil{CSIRO ICT Centre, University of Queensland}
DANIELA RUS
\affil{Massachusetts Institute of Technology}}





\begin{abstract}
 Long-term sensor network deployments demand careful power management.
While managing power requires understanding the amount of energy harvestable from the local environment, current solar prediction methods rely only on recent local history, which makes them susceptible to high variability.
In this paper, we present a model and algorithms for distributed solar current prediction, based on multiple linear regression to predict future solar current based on local, in-situ climatic and solar measurements.
These algorithms leverage spatial information from neighbors and adapt to the changing local conditions not captured by global climatic information.
We implement these algorithms on our Fleck platform and run a 7-week-long experiment validating our work.
In analyzing our results from this experiment, we determined that computing our model requires an increased energy expenditure of $4.5$mJ over simpler models (on the order of $10^{-7}\%$ of the harvested energy) to gain a prediction improvement of $39.7\%$.
\end{abstract}

\begin{document}

\maketitle

\begin{bottomstuff}
Author's address: E. Basha, Department of Electrical and Computer Engineering,
University of the Pacific, 3601 Pacific Avenue, Stockton, CA
95211.\newline
\end{bottomstuff}

\section{Introduction}
Resource-constrained sensor networks  require efficient management of their power usage to ensure long-term operation.  
Research on energy-efficient sensor networks has focused on  reducing the hardware's power needs and on developing models and policies in software to control system behavior.
Complementary to these areas is understanding the energy available to recharge the system.
This understanding informs the policies running on low-power hardware systems, helping create a smarter comprehensive energy management strategy.
The questions then arise of how much power is available from the environment to recharge the system and how much will be available in future days? We focus here on solar energy harvesting as a representative case.

Recent work on solar current prediction in sensor networks has focused on Persistence, where predictions assume a perfect correlation with current observations, and Exponentially Weighted Moving Average~\cite{hsuISPLED2006,kansalDAC2006}, which uses a window of recent measurements to predict future solar current. All of the existing prediction methods use only a node's local observation for prediction, which may include a high degree of variability. One solution to address the local variability is to base a node's prediction of solar current not only on its local measurements, but also on those of its neighbors. While individual node measurements have similar likelihoods for bias or noise, the joint consideration of multiple node measurements reduces the variability (following the central limit theorem~\cite{jaynes2003}). Furthermore, the environmental context that determines the amount of harvestable energy at sensor nodes, whether through sun, wind or water, is typically common to neighboring nodes, leading to a high degree of correlation in their conditions. The common environmental context further motivates the need for collective prediction.

\subsection{Solar Prediction Overview}
This paper presents a collective solar current prediction model that enables multiple nodes to participate in the prediction process.  Nodes share their solar and microclimate measurements with neighbors.
Collecting all these data and creating these time series then allows a node to locally predict its future available energy either centrally on its own processor or in a distributed manner with its one-hop neighbors.

Our solar prediction approach is based on multiple linear regression models.
These statistical models provide a powerful tool for predicting future time series while remaining simple enough to compute on a sensor network compared to other machine learning approaches.
Prediction involves three steps: (1) gathering an initial set of calibration data, (2) self-calibrating the model, and (3)  predicting the future solar current.
Steps (1) and (2) only run at startup and, after that, they run periodically  or whenever the prediction error exceeds some metrics in order to ensure adaptation to changing conditions.
In the rest of the paper,  we refer to step (3) as MLR prediction to distinguish it from the overall prediction algorithm.

\begin{figure}
\begin{center}
\includegraphics[width=0.4\columnwidth]{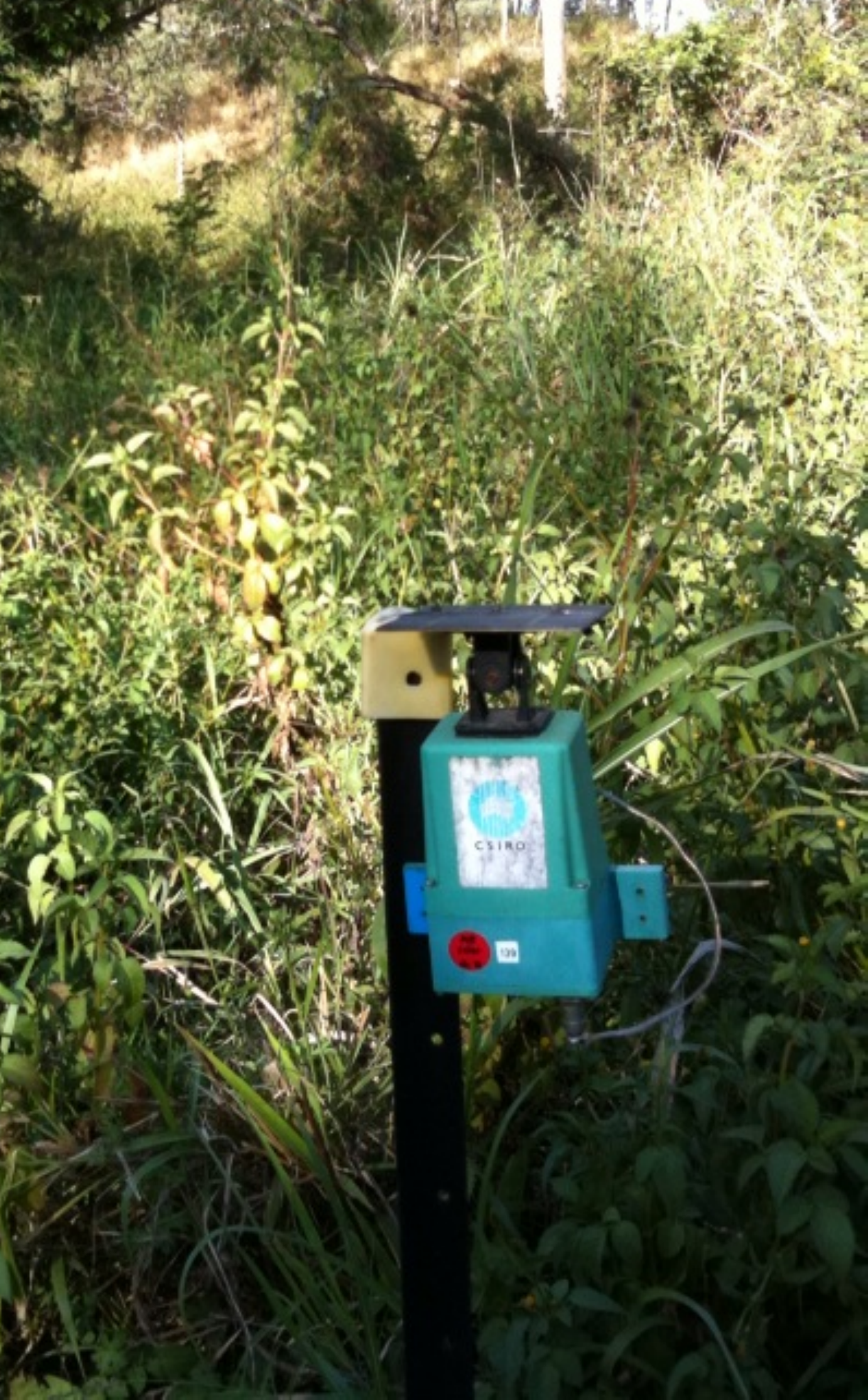}
\end{center}
\vspace{-0.5cm}
\caption{Fleck Node Installed on Campus}
\vspace{-0.2cm}
\label{fig:node}
\end{figure}

Collective prediction involves a communication overhead relative to local prediction, as nodes need to exchange their observations regularly. Centralizing the prediction operations at one node would provide the simplest implementation. However, as the network size grows, the communication overhead for transferring readings from all nodes to a single node becomes prohibitive. Additionally, the storage and computational constraints for data from hundreds of nodes may also become an issue. Distributed prediction within a local neighborhood alleviates the scalability issue and provides fault tolerance in the failure of the single point. Because each node performs its computation locally, it is more responsive to fine-grained changes in the environment. We thus focus our work on distributed prediction. 

This paper shows the feasibility of a distributed solar prediction strategy by developing distributed algorithms capable of computing these models and by implementing the models on a sensor network.
The core component of the distributed algorithms is a distributed pseudoinverse algorithm. To the best of the authors' knowledge, this paper is the first to propose and develop a distributed version for sensor networks, which forms a key contribution of our work.

Our models use solar current, humidity, soil moisture, air temperature, and wind speed to predict future daily average solar current.
With over 20 months of data from an installation in Springbrook, Australia, we improve over  previous models by up to 20\% using environmental data  or spatial data.
We also implement these models on the Fleck platform as shown in Figure~\ref{fig:node} and demonstrate their functionality during a 7-week-long test.
With the data from this test, we analyze the energy usage of our algorithms, determining they require $4.3\times10^{-7}\%$ of the weekly energy gathered by the system while providing a $39.7\%$  reduction in root mean squared error~(RMSE) over EWMA models and a $63.9\%$ improvement over Persistence.

\subsection{Contributions and Paper Organization}
The contributions of this paper are:
\begin{itemize}
\item Proposal of a distributed solar prediction model in resource-constrained devices: Section~\ref{sec:solar-model} describes the justification behind our model choice and the multiple linear model for solving this problem.
\item Design of distributed multiple linear regression (MLR) algorithms for predicting daily solar current on resource-constrained devices: Section~\ref{sec:solar-algs} motivates the distributed form, describes the solar current prediction algorithm, outlines the calibration algorithm, details each of the three sub-steps of the calibration algorithm, and analyzes the impacts of distributing the algorithm compared to computing it centrally.
\item Validation of the algorithms' accuracy gains through simulations on several months of empirical data: Section~\ref{sec:solar-test} provides three simulation data sets and provides results from those data sets, including an analysis of the effects of spatial and temporal data on the MLR algorithm.
\item Empirical evaluation of the algorithms on a small testbed of sensor nodes: Section~\ref{sec:solar-implement} explains the hardware platform, discusses the implementation of the algorithms on that platform, describes two field experiments and their results, and analyzes the energy impact of the prediction method.
\end{itemize}

In addition to the contributions, the paper describes related work in Section~\ref{sec:solar-related}, discusses some insights and future work in Section~\ref{sec:solar-discuss}, and concludes with Section~\ref{sec:solar-conc}.

\section{Related Work} \label{sec:solar-related}
Our work relates to two different areas: energy prediction in sensor networks and distributed regression in sensor networks. \newline

{\bf Energy Prediction}\newline 
Past research projects into energy prediction focus on predicting the future harvestable energy to input into power management models.
\cite{moserDAT2008} perform offline linear programming techniques to predict future energy in addition to control methods; the paper describes a simulation of this with no field instantiation.
\cite{Lu:2012:SFP:2185677.2185738} compute offline predictions of sunlight for daylight harvesting.
Their approach combines regression analysis with similarity identification and ensemble predictions; the work focuses on finer-grained predictions than ours in addition to its online focus.
\cite{5508260} include National Weather Service forecasts in an offline model to predict solar panel output; as described in their paper, the work is specific to their solar panel system with a focus on data analysis and offline simulation.

\cite{hsuISPLED2006} and \cite{kansalDAC2006} examine a prediction method using an Exponentially Weighted Moving Average Model; we discuss this approach in our simulation results.
Extensions to EWMA expand the set of data included in the moving average computation as described in \cite{5934952}, \cite{5601116}, and \cite{5172412}. 
\cite{Bergonzini2010766} explore EWMA, expansions to EWMA, and additional models, including neural networks.

Our work also provides a energy harvesting prediciton method.
A key difference between these approaches and our work is their focus on hourly predictions; expanding their models to daily predictions of the form we attempt in this paper either requires multiple years of data or reverts to a form of EWMA that we discuss in Section~\ref{sec:simulation}. 
Additionally, our work utilizes a more complex, richer set of input data and provides in-situ, local predictions.
Unlike previous models, our prediction model is based on multiple linear regression and can readily support any combination of environmental variables and spatial neighbor data to forecast available solar energy.\newline

{\bf Distributed Regression}\newline
Only a small amount of research exists in performing distributed regression on a sensor network.
\cite{guestrinIPSN2004} provide a distributed regression algorithm to help reduce the amount of data the network needs to communicate while allowing reasonable reconstruction of the node measurements back at the source.
Their approach assumes sparse matrices and utilizes kernel linear regression, a special case of linear regression.
In implementing kernel linear regression, they use Gaussian elimination to compute the weighting parameters.
Our approach uses a different, orthogonal method of computation; this method decreases the limitations on the matrix structure and removes the numerical issues seen with Gaussian elimination~\cite{golubMatrixBook1996}.

Our work here describes algorithms of similar flavor as in ~\cite{BashaSenSys2008} in the context of a different problem.
We present a fully decentralized solution as compared to the centralized results in ~\cite{BashaSenSys2008}, which suggested distributed solutions as future work, but focused on validating the use of regression models for river flood prediction and developing a sensor network system to support it.

Key in the development of distributed solutions is the distributed pseudoinverse.
This calculation does exist on multi-processor systems; examples include \cite{bensonBook1986,golubMatrixBook1996,milovanovicMCM1992,panPAA1990}.
Sensor networks differ from standard multi-processor work in several regards: lower processing capability, smaller memory, and wireless communication where broadcasting to all is easier than point-to-point communication with each node.
These differences require innovations in the computation of the distributed pseudoinverse.
To our knowledge, no prior work develops a distributed pseudoinverse for wireless sensor networks; this is a useful contribution of our work to the general sensor network community.
Aspects of this work were introduced and discussed in \cite{bashaPhD2010}.

\section{Prediction Model} \label{sec:solar-model}
In this section we describe a statistical model for predicting future daily average solar current using local measurements from a spatially-distributed sensor network.
This prediction can provide a key input into energy management of sensor networks, enabling intelligent policies and efficient usage of this resource.

\subsection{Model Justification}
Statistical models utilize in-situ measurements and a recorded time history of these measurements to develop spatially-distributed models.
By using only the observed record of data, these models can self-calibrate and morph to accommodate changes within the network, the environment, and other factors.
This makes them ideal for the purposes of sensor networks where deployments are dynamic, the surrounding environment may not be fully understood, and autonomy of the network from human intervention is desired.

Within statistical models, we choose to focus on the class of linear models, specifically multiple linear regression.
These models provide strong predictions within a computational framework that a sensor network can compute.
Sensor networks have limited computational and energy resources; defining a model that has the option to run on the network provides support for direct use of the prediction on the nodes for energy management.
MLR models allow for this in-network computation, unlike other machine learning approaches.
Additionally, the models lend themselves to automated calibration approaches that are computationally tractable as well as online and incremental computation.
Finally, the model structure adapts to the data types and ranges available, another benefit when considering rapidly scaling sensor networks.

\begin{algorithm}[!b]
\caption{Solar Current Prediction Model}\label{alg:pred}
\begin{algorithmic}[1]
\State $\phi:$ past daily average solar current
\State $\theta:$ vector of other nodes solar current 
\State $\rho:$ vector of other environmental measurements
\State $N:$ \# past solar current values used
\State $Q:$ \# other solar current values used
\State $P:$ \# environmental values used
\State $b:$ predicted daily average solar current
\State $e:$ prediction error
\State $T_{T}:$ training time window
\State $T_{L}:$ prediction lead time
\State $T_{R}:$ recalibration time window\\
\State $T_{TL} = T_{T}-T_{L};$\\

\Comment{Compute initial coefficients and prediction}
\State $\phi_N \gets [\phi(1:T_{TL}-N), .., \phi(1+N:T_{TL})]$ \label{alg:setupline1}
\State $\theta_P \gets [\theta(1:T_{TL}-P), .., \theta(1+P:T_{TL})]$
\State $\rho_Q \gets [\rho(1:T_{TL}-Q), .., \rho(1+Q:T_{TL})]$
\State $X \gets [\phi_N, \theta_P, \rho_Q]$ \label{alg:setupline2}
\State $C = ((X * X^T)^{-1} * X^T) * b(1+T_{L}:T_{T})$ \label{alg:calibration1}
\State $b(1+T_{L}:T_{T}) = X * C$ \label{alg:incerror1} \\ 

\Comment{Recompute using prediction error}
\State $e = b(1+T_{L}:T_{T}) - \phi(1:T_{T}-T_{L})$ 
\State $X \gets [\phi_N, e, \theta_P, \rho_Q]$ \label{alg:setupline3}
\State $C = ((X * X^T)^{-1} * X^T) * b(1+T_{L}:T_{T})$ \label{alg:calibration2} \\

\For{$t=T_{T}+1$ to $...$} \label{alg:loop}
\Comment{Predict}

\If{$(t \% T_R) == 0$} \label{alg:recal} \\
\Comment{Recalibrate coefficients}
\State $e = b(t-T_{T}:t) - \phi(t-T_{T}-T_{L}:t-T_{L})$
\State $\phi_N \gets [\phi(t-T_{TL}:t-N), .., \phi(t-T_{TL}+N:t)]$
\State $\theta_P \gets [\theta(t-T_{TL}:t-P), .., \theta(t-T_{TL}+P:t)]$
\State $\rho_Q \gets [\rho(t-T_{TL}:t-Q), .., \rho(t-T_{TL}+Q:t)]$
\State $X \gets [\phi_N, e, \theta_P, \rho_Q]$
\State $C = ((X * X^T)^{-1} * X^T) * b(t-T_{T}:t)$ \label{alg:calibration3}
\EndIf\\
\Comment{Compute prediction}
\State $e = b(t-T_{L}) - \phi(t)$ \label{alg:error}
\State $\phi_N \gets [\phi(t-N), .., \phi(t)]$
\State $\theta_P \gets [\theta(t-P), .., \theta(t)]$
\State $\rho_Q \gets [\rho(t-Q), .., \rho(t)]$
\State $X \gets [\phi_N, e, \theta_P, \rho_Q]$
\State $b(t+T_{L}) = X * C$ \label{alg:prediction2}
\label{alg:pred:model}
\EndFor
\end{algorithmic}
\end{algorithm}

\subsection{MLR Model}
The multiple linear regression  model linearly combines $N$ weighted past observations of all relevant variables at time $t$ to predict a future variable of interest at time $t+T_L$.
To determine the weighting factors, the model looks at a past history of these variables, defining a training set of time $T_T$ which outlines a matrix of these past observations.
Calibration then consists of inverting this matrix and multiplying it by the actual observations of the prediction variable.
As an aside, to place our models in the context of the widely known autoregression models (such as AR, MA, and ARIMA), regression models form the larger class of which AR models are a specialized form.
AR models focus on predicting outputs of shock-driven systems, where the variables used for the prediction directly relate, or are errors related to, the variable being predicted~\cite{box1976}.
This relationship leads to the ``auto'' descriptor of these models and the weighting parameters used to define the relationship have definitions related to the autocorrelations of the process.
These models do not allow the use of other variables; once we include measurements other than a single node's solar current, we need to use multiple linear regression (MLR) models.

Using MLR models, we wish to predict:
\begin{equation}
  b=f(\phi,\theta,\rho)
\end{equation}
where $b$ is the future daily average solar current.
Variable $b$ is a function of $\phi$, a time history of daily average solar current; $\theta$, a time history of our neighbors' daily average solar current; and $\rho$, a time history of the node's other environmental measurements such as humidity, air temperature, leaf wetness, wind speed, and other values.
These latter two sets of variables help outline the external factors affecting the amount of energy harvestable by solar including weather and seasonal conditions.
The regression model provides flexibility in defining these sets, allowing the sets to reflect the variables available in the network.

Algorithm \ref{alg:pred} outlines a multiple linear regression model for predicting solar current using past solar current, nearby neighbors' solar current, and any available environmental variables.
In the algorithm, Lines \ref{alg:setupline1} through \ref{alg:setupline2} setup the calibration matrix.
For each variable, we can use any number of past values to define the linear prediction; in calibration each past value becomes a column within the matrix.
For example, in using the past solar current observations, we could only use that occurring at $t$, or we could add $t-1$, $t-2$, etc. as we design the actual implementation of the model.
In our algorithm description, we define the number of past values that the model uses as $N$, $Q$, and $P$ for the node's solar current ($\phi$), the neighbors' solar current ($\theta$), and the environmental measurements ($\rho$), respectively.
We load the matrix, $X$, with this data set over the calibration window defined.
Line \ref{alg:calibration1} performs the calibration step and generates a coefficient vector, $C$.
To include the prediction error in the model, we then predict over the training window we just used by multiplying our calibration matrix by our coefficients (Line \ref{alg:incerror1}).
We subtract the observed record from this prediction, thus generating our error, and include this in our new calibration matrix of Line \ref{alg:setupline3}.
Recomputing the coefficients based on the new calibration matrix occurs in Line \ref{alg:calibration2} and, with these coefficients, we begin predicting the future solar current in Line \ref{alg:loop}.
This loop continues forever, recalibrating after a full recalibration time window passes (Line \ref{alg:recal}), computing the prediction error based on the latest observation (Line \ref{alg:error}), and predicting future solar current (Line \ref{alg:prediction2}).

Algorithm \ref{alg:pred} describes the overall model and a centralized version of the operations.
We next can consider how to distribute this among the nodes, especially the storage of the data within the network and the complex calibrations of Lines \ref{alg:calibration1}, \ref{alg:calibration2}, and \ref{alg:calibration3}.

\section{Distributed Algorithms for Solar Current Prediction} \label{sec:solar-algs}
To implement this model, we divide the problem into two parts: prediction and calibration.
Prediction computes the future daily average solar current and, to avoid confusion with the general goal of prediction, we will refer to this as MLR prediction since that is our method.
This computation occurs at regular intervals for the duration of the network operation.
Calibration computes the coefficients necessary to perform MLR prediction based on a data set of past measured values.
Calibration occurs once this data set exists within the network and only sporadically after that.
We develop distributed algorithms to perform both computations within the sensor network.

Note that these algorithms operate on top of the existing communication protocols of the network.
The algorithms utilize the existing ability to route packets to other nodes and efficiently combine broadcast messages from a range of protocols similar to methods described in~\cite{HansenIPSN2011}.
Thus any node can reach any other node without additional overhead or operation in our algorithms.

This section provides a motivation for distributing the prediction, a description of the distributed prediction algorithm, an outline of the distributed calibration algorithm including the sub-algorithms, and an analysis of the computation and memory requirements.

\subsection{Motivation for Distributed Prediction}
Our goals in performing these in-situ models are to optimize communication of the model (thus saving energy), optimize computation, and optimize the usefulness of the prediction.
Achieving these goals requires distributed algorithms and shared storage of the matrix.
A useful prediction model requires incorporating a wide variety of data into the matrix, possibly more data than a single node can store.
To store the matrix, we need to distribute the data among the network, which provides a secondary benefit by ensuring the prediction is more robust to node failures.
Because we have to distribute the data matrix storage, we can distribute the computation also, sharing the computational load and reducing the communication load.
This achieves all our goals and motivates our distributed algorithms.

A common argument against distributed computation is the existence of a base station or gateway node of higher computational power nearby, or the nearby presence of a central location to provide the computation.
Some deployments do fit these scenarios with systems located on campuses or within communication reach via high-powered gateways.
However, some deployments and applications do not, especially in the area of environmental monitoring where we focus.

Intermittent connectivity with a sensor network gateway may preclude centralized solar prediction. Distributed prediction, on the other hand, supports delay-tolerant sensor networks and networks with mobile gateways. Consider a group of static nodes that are disconnected from their gateway for prolonged periods (such as when an Unmanned Aerial Vehicle acts as a mobile gateway), but still need to sample and log sensor data locally for later uploading. These nodes would need to locally predict available solar energy among themselves. A distributed prediction approach thus enables the nodes to act autonomously and independently of their connectivity to a base station. 

Other logistical reasons also favor distributed prediction. Deployments in remote areas may not have reasonable communication range back to a central office or may not allow for the power necessary to run a higher power gateway (lugging batteries into forests is not always feasible).
Additionally, these deployments may utilize the same processing unit for the gateway as the rest of the network.
Designers may choose this latter design option as it allows redefinition of who is the gateway, reduces gateway power requirements, and simplifies the system.
However, it then provides no computational benefit for centralized algorithms.

Where gateways do exist that could provide centralized computation, issues of scalability and fault tolerance arise.
At some point, the number of nodes needing individual predictions will exceed even the computational capability of the gateway and/or the communication requirements to reach all the nodes will exceed those needed to perform local distributed computations.

Additionally, the gateway will fail; it is just another node, albeit a more powerful one.
When this node fails, repair may take several days, depending on location, while the system flounders without the centralized control the gateway provided.
In the case of energy management, the system could follow policies that lead to an overall decline in performance resulting in a shorter system lifetime, or, more drastically, the failure of individual nodes.

All of the above reasons confirm  the usefulness of distributed algorithms, especially to support the overall general case and allow deployment of sensor networks wherever we want, not limited by the need for gateway communication, easy access, or centralized control.

\subsection{MLR Prediction}
We compute the MLR prediction of the daily average solar current as a linear combination of scaled variables.
This computation could occur centrally (depending on the number of variables).
Distributing it shares the computation with limited additional costs as the network already communicates measurements and status; we can simply add the scaled variables to these pre-existing messages with only a small number of byte transmission costs incurred.
We thus compute:
\begin{equation}
  b_{t+L}=x_{t0}a_{t0}+x_{t1}a_{t1}+...+x_{tn}a_{tn}
\end{equation}
in a distributed fashion using Algorithm \ref{alg:distPred}.
In this format, $a$ is a vector containing a node's variables at time $t$ consisting of solar current, that node's environmental measurements, and neighbors' solar current.
The sub-index $j$ indicates which variable as all $a_{tj}$ values are measured at the same time $t$.
Variable $x$ is a vector of the weighting parameters, provided by our distributed calibration algorithm.
Variable $b$ is our prediction.

\begin{algorithm}
\caption{Distributed MLR Prediction}\label{alg:distPred}
\begin{algorithmic}[1]
\For{Each node $i$}
\State Measure $j$ values of $a_{t}$
\State $\tau_i=a_{tj}x_{tj}$
\State Transmit $\tau_i$ \\
\State Receive from all other nodes $\tau_{k \ne i}$
\State $b_{t+L}=\tau_i+\tau_{k \ne i}$
\EndFor
\end{algorithmic}
\end{algorithm}

In Algorithm \ref{alg:distPred}, each node measures some portion, $j$, of the $a$ values at current time, $t$.
Here we allow nodes to maintain more than one variable of the computation and, while $j$ suggests that each stores the same number of variables, all of the algorithms allow for an unequal number of variables stored by each node (in cases where this reduces communication for example).
The node then multiplies these $a$ values by its portion of the stored $x$ values and communicates the result to the other nodes.
Each node, upon receiving the prediction components from the other nodes, adds the other components to its own MLR prediction component, thus computing its own prediction $L$ time intervals in the future, $b_{t+L}$.

Nodes have two options for determining a prediction of its own solar current.
First, assuming the solar current of one node correlates well with the other nodes (such as might be the case if all are placed with full solar exposure), all nodes can collaborate on predicting the solar current of one node, computing the MLR prediction value, $b_{t+L}$.
All participating nodes then use that value as their prediction.
To achieve this, each node maintains some portion of the data and the associated weighting coefficients.
In a centralized MLR prediction, nodes would have to communicate data values to the central node anyway; this distributed form communicates aggregated weighted values instead, reducing the number of values to communicate and sharing the computation.

Or, if the similarity in solar current does not exist, the nodes each compute their own solar current prediction based on their own solar current measurements.
Distribution occurs when including neighbors' solar current measurements and environmental variables; the latter should correlate well among the nodes allowing all to share the same measurements.
With each node performing a MLR prediction, a node already maintains its solar current values and just needs to store additional weighting coefficients for its neighbors.
Sharing environmental variables allows nodes to divide them equally among the network, storing some portion and associated weighting coefficients.
The increase in precision due to the increase in data available to the model offsets the potential increase in communication (potential as the number of values communicated is small and the system can append them to existing messages).

The first option reduces computation and communication while generating a reasonable solar current prediction to input into power management systems; the second improves each node's prediction with an increased cost in computation and communication, and decrease in scalability.
Choosing between the two depends on the structure of the network, similarities between node placements, and requirements for precision in the solar current predictions.
For simulation validation, we only focus on a single node running its own MLR prediction based on its own historical measurements and on historical measurements from its neighbors; for field experiments, the system uses the first method with a single node aggregating the MLR prediction for everyone.

\begin{figure}[h]
\begin{center}
\includegraphics[width=0.65\columnwidth]{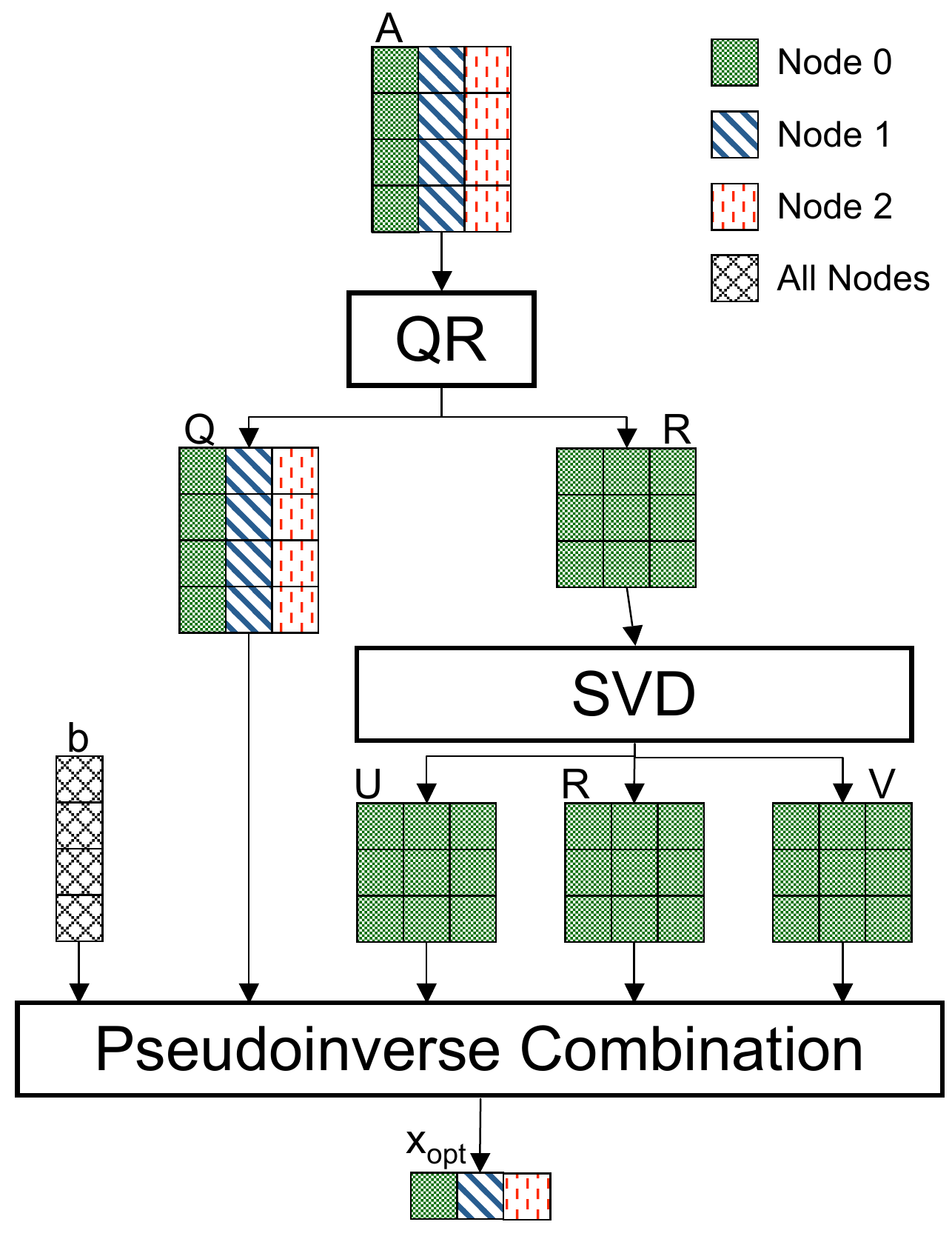}
\end{center}
\caption{Flow of Calibration Algorithm}
\label{fig:algFlow}
\end{figure}

\subsection{Calibration}
The calibration step provides the weighting coefficients by computing:
\begin{equation}
  X=((AA^T)^{-1}A^T)b
\end{equation}
where $A$ is the matrix of historical measurements of solar current and environmental conditions, $b$ is the vector of past observed solar current $L$ time steps after the last solar current values in $A$, and $X$ is the computed coefficients.
$((AA^T)^{-1}A^T)$ is the standard Moore-Penrose pseudoinverse of a real matrix, which also includes a computationally intensive square matrix inverse and three matrix multiplication steps \cite{golubMatrixBook1996}.
To perform this computation on a sensor network centrally is infeasible and undesirable as $A$ is usually too large to store on one sensor node (much less all the additional storage necessary for the pseudoinverse computation); therefore we need a distributed algorithm where each node maintains some number of columns.
As no distributed pseudoinverse exists for sensor networks, we need to innovate on existing algorithms.

We break the problem into three steps as shown in Algorithm \ref{alg:distCal}: (1) a QR decomposition to reduce issues related to the rectangular nature of $A$, (2) a singular value decomposition (SVD) to compute the necessary pieces, and (3) a final pseudoinverse step to combine all the pieces into a solution.
We discuss each of these algorithmic steps in detail.

\begin{algorithm}
\caption{Distributed Calibration}\label{alg:distCal}
\begin{algorithmic}
\State $[Q, R] = qr(A)$
\Comment Dist. QR Decomposition: Alg. \ref{alg:qr}
\State $[U,R,V] = svd(R)$
\State $x_{opt}=VR^{-1}(QU)^Tb$
\Comment Pseudoinverse: Alg. \ref{alg:picalregCent}
\end{algorithmic}
\end{algorithm}

{\bf Distributed QR Decomposition}\newline
The QR decomposition requires as input any matrix $A$, which is of size $m \times n$ where $m \geq n$.
It then decomposes $A$ into $Q$ of size $m \times n$ and $R$ of size $n \times n$.
This decomposition allows for squaring off and shrinking the matrix used by the SVD, a key component of the pseudoinverse.

We compute the QR using a modified Gram-Schmidt form as outlined in Algorithm \ref{alg:qr} \cite{golubMatrixBook1996}.
This form decomposes the matrix in column order allowing us to store some number of columns per node.
To ease explanations, we assign an equal number of columns to each node, specifically $p$ columns per node.

At the beginning of the algorithm, each node has $p$ columns of $A$, consisting of $m$ past data values already gathered by the node.
We perform the computation in place, replacing the columns of $A$ with columns of $Q$.
For additional storage, the computation will need $np$ data values stored for $R$ and $m$ temporary values.

The algorithm begins with Node 0, which is the node storing column 0.
This node computes the $l_2$ norm of the $Q_0$ column ($||Q_0||_2 = \sqrt{\sum_{i=0}^{m}{Q_0(i)^2}}$ ) to fill the $R_{00}$ entry and then divides the $Q_0$ column by that value to obtain the decomposition final value for $Q_{0}$ column.
It then communicates the $Q_0$ column to all other nodes.
Each node uses the data to update its own columns.
If Node 0 stores more columns ($p>1$), it retains control and computes final values for both $R_{11}$ and $Q_1$, communicating $Q_1$ to the other nodes.
Control switches to Node 1 when the algorithm reaches column $p$.
As it finalizes its columns, it communicates them and the algorithm continues, switching control after every $p$ columns.
The algorithm terminates once all columns have been computed, resulting in a $Q$ matrix of size $m \times n$ and a $R$ matrix of size $n \times n$, with $Q$ completely replacing $A$.
Each node maintains $p$ columns of $Q$ and $R$.
By performing the QR decomposition, we can focus our remaining operations on the much smaller, square $R$ matrix, simplifying communication and reducing concerns about the structure of the matrix (condition, linear independence, etc.).

\begin{algorithm}
\caption{QR Decomposition}\label{alg:qr}
\begin{algorithmic}
\State $Q \gets A$
\State $R \gets 0$\\

\State Node 0 begins:
\State $ \ \ R_{00}=||Q_0||_2=\sqrt{\sum_{i=0}^{m}{Q_0(i)^2}}$
\State $ \ \ Q_0=Q_0/R_{00}$
\State $ \ \ $Communicate $Q_0$ to other nodes \\

\For{$Node=0:n/p-1$}
\State Node receives $Q_i$ column
\For{$k=p(Node):p(Node+1)-1$}
\State $R_{ik} = Q_i^TQ_k$
\State $Q_k=Q_k-R_{ik}Q_i$
\If{$(i+1==p(Node)) \ or \ (i==k)$}
\State $R_{kk}=||Q_k||_2=\sqrt{\sum_{i=0}^{m}{Q_k(i)^2}}$
\State $Q_k=Q_k/R_{kk}$
\State Communicate $Q_k$ to other nodes
\EndIf
\EndFor
\EndFor
\end{algorithmic}
\end{algorithm}

{\bf Singular Value Decomposition}\newline
The Singular Value Decomposition (SVD) uses $R$, the smaller matrix, to generate the components of the pseudoinverse: $U$, $R$, and $V$; each is a $n \times n$ matrix.
Because, for this application, $n$ is small, we compute this portion centrally and reduce communication overhead; we do need to first collect the data at a central node as $R$ is distributed among the system following the QR decomposition.

We compute the SVD based on a cyclic Jacobi procedure for mesh processors outlined by Brent~\etal~\cite{brentJVCS1985}.
Because this step occurs after the QR decomposition, we ensure the input matrix, $R$, is square with dimensions $n \times n$.
If $n$ is odd, the algorithm pads the matrix with a column and row of zeros to make it even.

This algorithm works on a block level with the smallest block being a $2$ by $2$ matrix.
The node computes a diagonal block of the matrix and then normalizes the values to properly order the singular values in $R$.
Finally, these values scale the non-diagonal blocks of the columns, the $U$ matrix, and the $V$ matrix.
Once these computations complete, the node rotates the data values and continues the computation until the rotation finishes and the stopping condition is reached.

{\bf Distributed Pseudoinverse Combination}\newline
Finally, once we complete the SVD, we combine the various sub-matrices to compute the coefficients, which consists of computing $A^+=VR^{-1}(QU)^T$ and then multiplying by $b$.
Algorithm \ref{alg:picalregCent} describes how we perform this computation in the case where $n$ is small enough to allow a centralized computation of the SVD.

We can split the computation into a centralized portion computed by Node 0 and a distributed portion computed by all nodes.
Node 0 computed the SVD so has matrices $U$, $V$, and $R$ (which is diagonal so only $n$ values).
It first computes $VR^{-1}U^T$, resulting in a $n \times n$ matrix.
Knowing the final multiplication by $b$ allows us to optimize the computation by multiplying $Q^T$ by $b$ before communicating these values.
Each node has $p$ columns of $Q$ and $p$ values of $b$; the transpose allows each node to compute $Q^Tb$ ($Qb$ would require storing rows and additional communication to reshuffle $Q$ after Algorithm \ref{alg:qr}).
Nodes then communicate their portion to Node 0, which performs the last multiplication step to compute the weighting coefficients.
The system completes calibration by communicating these values to those nodes participating in the distributed MLR prediction.

\begin{algorithm}[h]
\begin{algorithmic}
\caption{Pseudoinverse Combination for Linear Regression}\label{alg:picalregCent}
\State Central node computes the following:
\State $V=VR^{-1}$
\State $\mathit{T} =VU^T$ \\

\State Each node $i$ computes the following:
\State $q_i=Q_i^Tb_i$
\State Each node transmits $q_i$ with the central node storing the result vector of length $n$ \\

\State $x_{opt}=\mathit{T} q_i$
\end{algorithmic}
\end{algorithm}

\subsection{Algorithmic Analysis}
We now analyze our algorithms to determine the impact on computation, memory, and communication.
Table \ref{table:algAnalysis} shows the requirements of our distributed approach compared to a fully centralized model.
For the distributed case, we assume each node maintains one variable, which is the worst case scenario and results in $n$ nodes for the network. 
Therefore, in the table, $m$ defines the number of time samples for each variable while $n$ defines the number of variables as well as the number of nodes in the network.

For both computation and memory, the centralized approach reflects the requirements of the node performing the centralized computation.
The distributed case states the requirements for the controlling node, which must compute and store more than the other nodes.
Each explores the most impacted node; however, in the distributed case, the communication load is shared across all nodes while the centralized root node has the highest communication load of all the nodes.
In the centralized case, then, the leaf nodes can have a light communication load, which could lead to a longer overall system lifetime if the sink responsibility can move between nodes in the case of the original central sink node losing power.

On the communication side, in a centralized approach, all messages have to converge on the root node.
In converging on the root, the nodes communicate via a collection tree topology, which has a known communication pattern and known scaling limitations.
As the number of nodes grows, each packet needs to be transmitted and forwarded multiple times in a multi-hop collection tree with the number of transmissions dependent on the network topology.
We use the average case where the network topology is a binary tree for the centralized approach.
For the distributed approach, the nodes follow a daisy chain where one node initiates the prediction and passes the responsibility to the next node, requiring $2n$ iterations of broadcasting and waiting for replies.
As the table indicates, the communication load at the bottleneck node is equal between the two methods with both requiring $O(n)$ messages transmitted and received (although the total overhead may differ).
The centralized approach still needs to gather all the data while the distributed approach shares the computation and storage load, requiring intermediate communication for the model without the communication of all the raw data.

\begin{table}
\centering
\tbl{Analysis of Distributed and Centralized Algorithms - Results are for the Most Impacted Node in Both Cases\label{table:algAnalysis}}{
\begin{tabular}{|c|c|c|c|c|}
\hline
{} & \multicolumn{2}{|c|}{\bf Calibration} & \multicolumn{2}{|c|}{\bf MLR Prediction} \\ \hline
{} & {\bf Centralized} & {\bf Distributed} & {\bf Centralized} & {\bf Distributed} \\ \hline
Computation & $O(m^2n^3+n^3+mn)$ & $O(m^2n^2+n^3+m)$ & $O(n)$ & $O(n)$\\ \hline
Memory & $O(mn+n^2+n^2)$ & $O(m+n+n^2)$ & $O(1)$ & $O(1)$\\ \hline
Communication & $O(n)$ & $O(n)$ & $O(1)$ & $O(1)$ \\ \hline
\end{tabular}}
\end{table}

\begin{table*}[b]
\centering
\tbl{Intuition of Analysis Results for Overall Algorithm\label{table:intuitionAnalysisOverall}}{
\begin{tabular}{|l|r|r|r|r|r|} \hline
& & \multicolumn{2}{|c|}{\bf Calibration } & \multicolumn{2}{|c|}{\bf MLR Prediction} \\ \hline
{\bf Algorithm} & {\bf Matrix Size ($m \times n$)}  & {\bf Computation (ops)} & {\bf Memory (values)} & {\bf Computation (ops)} & {\bf Memory (values)} \\ \hline
{\bf Distributed} & $10\times10$ & 8252 & 262 & 10 & 1 \\\cline{2-6}
{\bf Linear} & $100\times10$ & 456812 & 442 & 10 & 1 \\ \cline{2-6}
{\bf Regression} & $100\times20$ & 1914267 & 882 & 20 & 1 \\ \cline{2-6}
 & $1000\times10$ & 45037412 & 2242 & 10 & 1 \\ \hline
{\bf Semi-} & $10\times10$ & 305864 & 450 & 10 & 1\\ \cline{2-6}
{\bf Distributed} & $100\times10$ & 754424 & 630 & 10 & 1\\ \cline{2-6}
{\bf Linear} & $100\times20$ & 4248949 & 1860 & 20 & 1\\ \cline{2-6}
{\bf Regression}  & $1000\times10$ & 45335024 & 2430 & 10 & 1\\ \hline
{\bf Centralized} & $10\times10$ & 351250 & 910 & 29 & 1\\ \cline{2-6}
{\bf Linear} & $100\times10$ & 4869250 & 12610 & 29 & 1\\ \cline{2-6}
{\bf Regression} & $100\times20$ & 40609400 & 16420 & 39 & 1\\ \cline{2-6}
 & $1000\times10$ & 450999250 & 1020610 & 29 & 1\\ \hline
\end{tabular}
}
\end{table*}

In examining the table, we first analyze MLR prediction.
Here centralized and distributed approaches have similar results due to the constant factors.
The computation similarity occurs as the most impacted node is either the central one, which then has to compute the full MLR prediction for everyone, or a distributed node, which computes its own MLR prediction based on received messages.
For communication, the centralized node needs to send one message with the prediction while the distributed node sends one message with its weighted value.
If we assume the centralized computation occurs as the node receives measurements from the other nodes, the centralized node only stores the MLR prediction value.
The distributed node only needs to save its MLR prediction as well, which will begin as its own weighted value.
These results support our earlier discussion that distributed MLR prediction is not essential although it does provide some fault tolerance properties in that the result does not depend on only one node, and some local autonomy for nodes to continue to predict solar power in delay-tolerant networks.

Examining the calibration side of the table, we see that the distributed approach requires a lesser amount of computation and memory per node.
Both computation and memory are split into three components to reflect the requirements of each of the three sub-algorithms.
For these algorithms, $m$ needs to be greater or equal to $n$.
Therefore, both computation and memory are dominated by the $m^qn^p$ terms.
These terms are larger by a factor of $n$ for the centralized approach.
Table \ref{table:intuitionAnalysisOverall} provides some insight into what this means for the scalability of the approach.
Notice that we use the term ``semi-distributed'' to indicate that we are examining a version where the SVD step is centralized and all other steps are distributed; this version supports the hardware implementation more completely and Section~\ref{sec:solar-implement} will discuss it further.
We outline the computation in terms of operations without consideration for how fast the processor performs and the memory in terms of values without defining the representation of those values, providing a system independent view.
Centralized quickly grows beyond the memory bounds of most microcontrollers, making the processing and computation quite difficult for larger sensor networks.

Fundamentally, for the calibration, this type of computation is not scalable centrally and must be distributed to be computable on a sensor network.
The next question is how effective is the model is at the prediction of solar power.

\section{Testing: Simulation} \label{sec:solar-test}
In this section we describe our simulation testing in Matlab to verify the functionality of this model in predicting solar current.
We simulate model functionality in different seasons, using different environmental sensors, and including spatial features through neighbors' solar data.

\subsection{Simulation Data}\label{sec:simulation}
We suggest that multiple linear models can accurately predict solar current, but need to verify this claim.
We have two existing data sets from a rainforest deployment in Springbrook, Australia that provides the relevant parameters.
This rainforest is a subtropical location with a winter characterized by temperature variations but with little rainfall or cloud cover to impact solar energy harvesting and a summer characterized by little temperature variations but significant rainfall and cloud cover that does impact solar harvesting.

\begin{figure}
\centering
  \subfigure{\includegraphics[width=0.88\columnwidth]{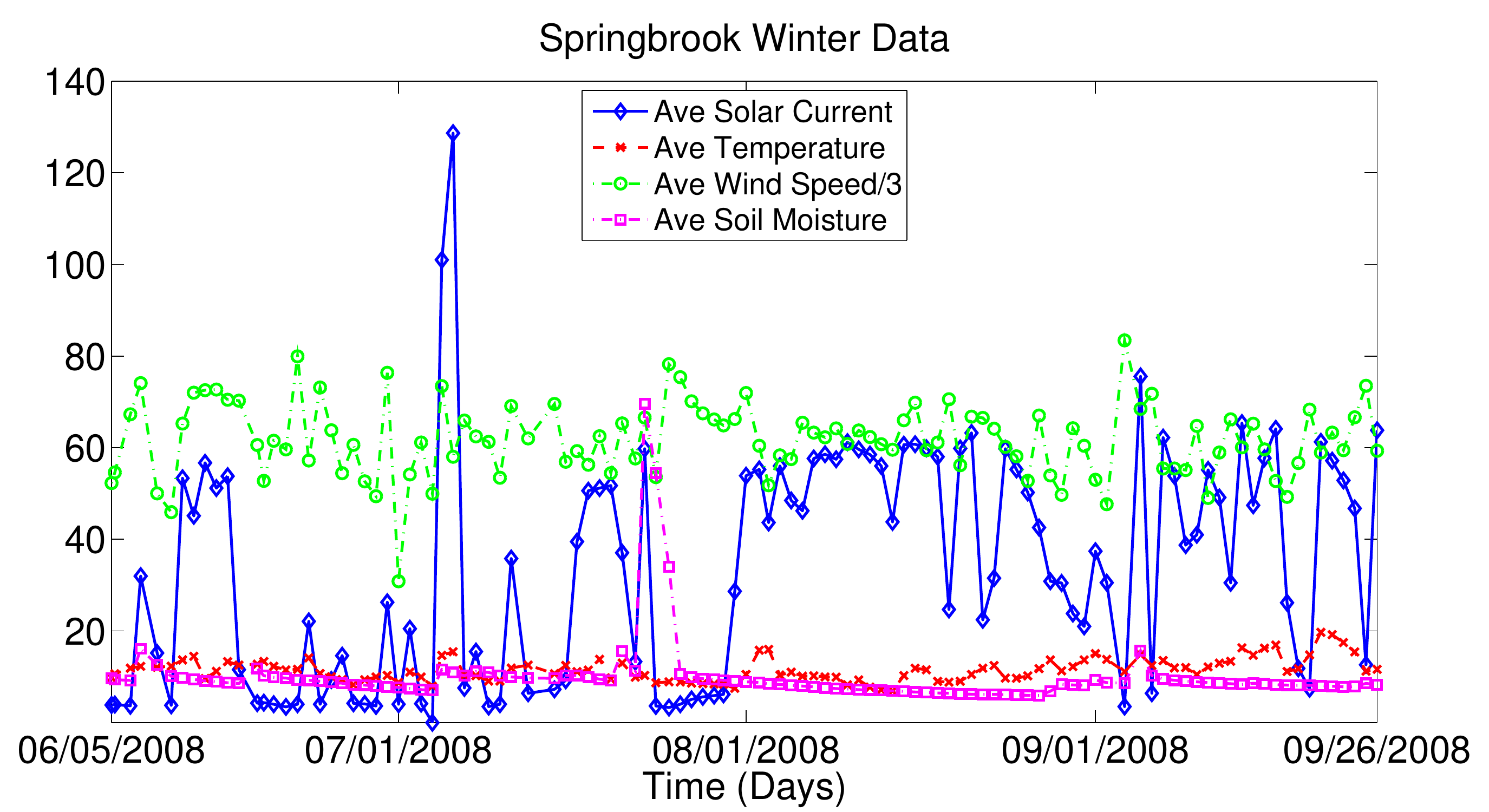}\label{fig:winterData}}\\
  \subfigure{\includegraphics[width=0.88\columnwidth]{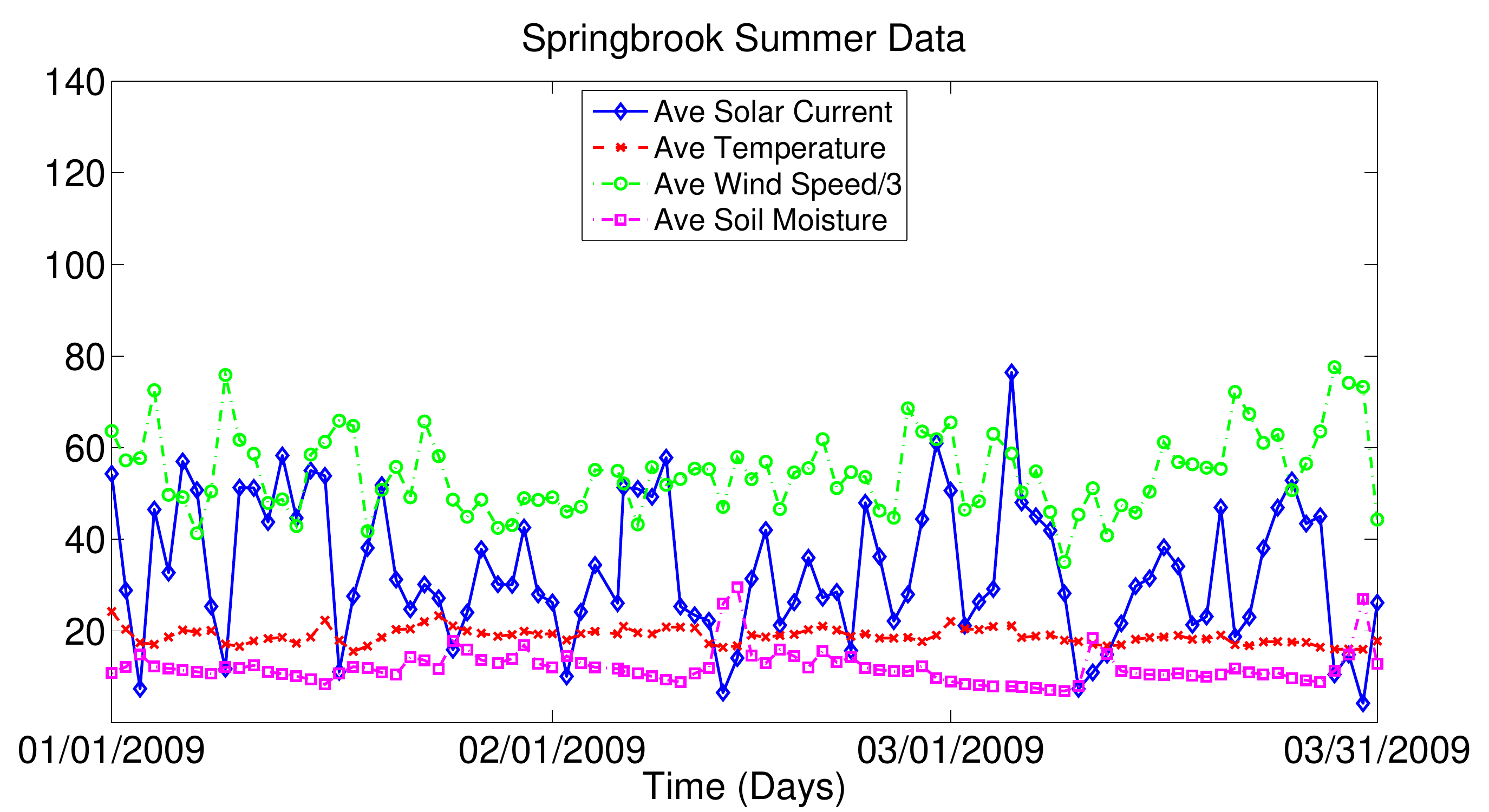}\label{fig:summerData}}\\
\caption{Springbrook Data Sets: (a) 2008 Winter Data and (b) 2009 Summer Data}
\label{fig:springbrook2008Data}
\end{figure}

\begin{figure}
\centering
  \includegraphics[width=0.88\columnwidth]{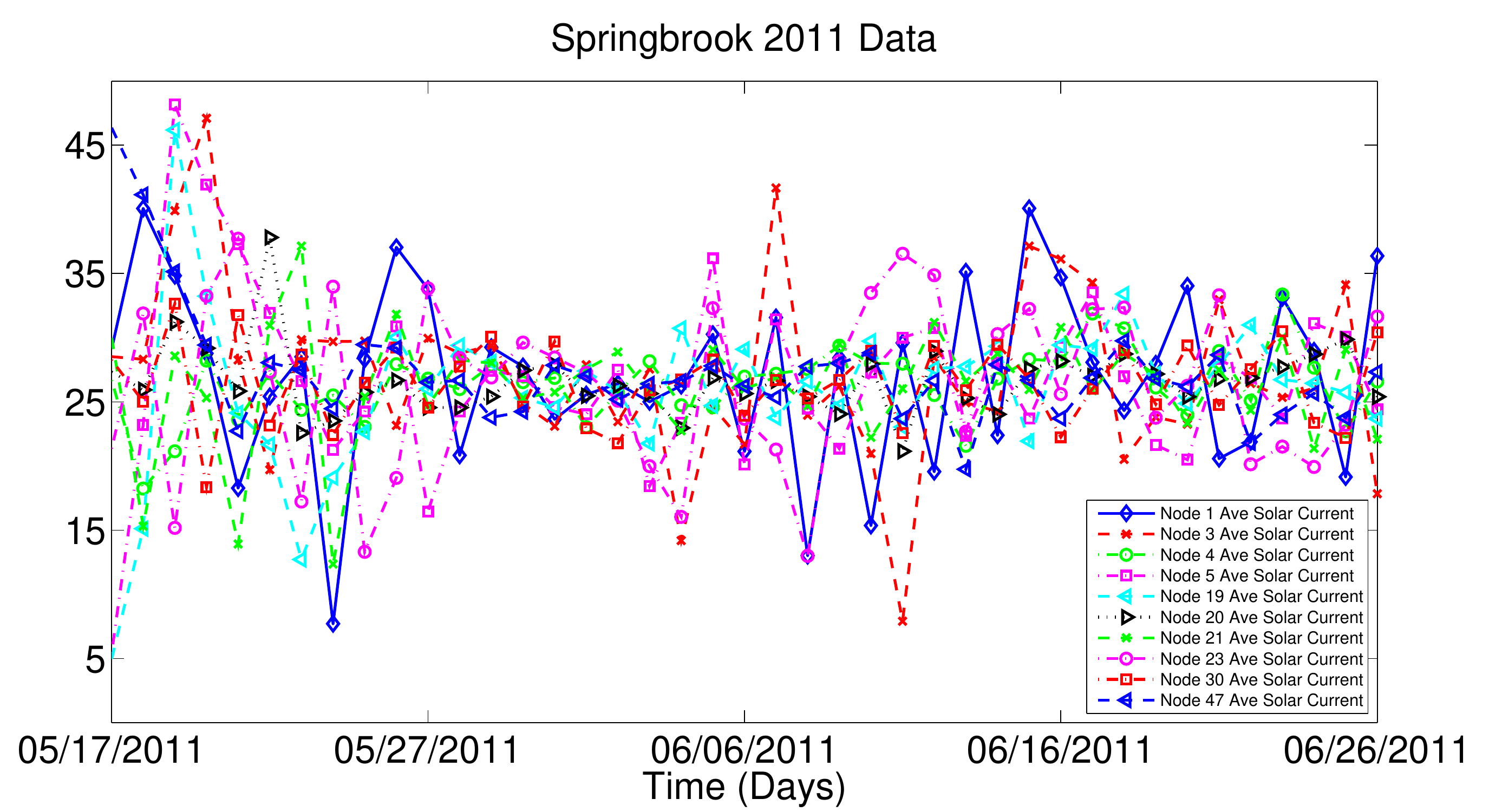}
\caption{Springbrook 2011 Fall Data}
\label{fig:springbrook2011Data}
\end{figure}

One data set consists of one year of solar current, humidity, soil moisture, air temperature, leaf wetness, wind speed, and wind direction, gathered by 10 sensor nodes.
Due to some small gaps in the operation of the network and gateway node, we split the first data set into a summer data set of three months from January through March and a winter data set of four months from June through September.
Our second data set occurs after the network size grew to over 50 nodes and consists of over two months of data in May through June of 2011.
Figures \ref{fig:springbrook2008Data} and \ref{fig:springbrook2011Data} display the measurements for these data sets.
Despite averaging them on daily boundaries, we still see a non-linear time series with no trends in solar current values from day to day.
There also appears to be no obvious correlation between the solar current and the temperature or the humidity in the 2008-2009  data.

\subsection{Testing and Validating Algorithms}
Next we implement Algorithm \ref{alg:distCal} in Matlab.
We predict the future daily average solar current two days in advance using seven days of data for calibration and training (a value chosen based on the amount of existing data, ideally this would increase as more data arrives), starting with the scenario where the model uses environmental variables and no neighbors' information.
To the measured values, we also add the three possibilities: (1) recalibrating when the error exceeds a threshold and at least four days of new data exist in the matrix, (2) including the prediction error in the calibration matrix, and (3) including the first derivative of the solar current in the calibration matrix.
We do not know which other variables most correlate to the solar current so vary all parameters and run the model over each possibility.
To determine which combination provides the best prediction, we evaluate the predicted time series using the root mean square error (RMSE) between the predicted and observed as well as the largest absolute error value.
Table \ref{table:environResults} outlines the combinations with the best results over both metrics.

To understand the bounds and convergence of the model, we compute the mean of the error between observed and predicted (also known as the mean residual) and the $95\%$ confidence interval around this residual.
The confidence interval provides a probabilistic bound on how much the residual will vary from the mean, allowing us to provide probabilistic bounds on the error and the convergence of the model.
Equation~\ref{eq:confInt} outlines the computation of the confidence interval~\cite{Ramsay2002StatisticalSlueth}.

\begin{equation}\label{eq:confInt}
  ConfidenceInterval = \frac{t_{1-c}/2*s}{sqrt(n)}
\end{equation}
where $s$ is the sample standard deviation, $n$ is the sample size, $c$ is the confidence interval, and $t_{1-c}$ is determined from established lookup tables~\cite{Ramsay2002StatisticalSlueth}.
Table~\ref{table:environResults} shows these values in the last two columns for all models.

In addition to the metrics, we compare our results to two other locally computable methods: persistence and exponentially weighted moving average (as suggested in \cite{hsuISPLED2006,kansalDAC2006}).
Persistence predicts that nothing will change and the future solar current value will be equal to the current solar current value.
Exponentially Weighted Moving Average (EWMA) computes \mbox{$b_{t+L}=\alpha b_{t+L-1}+(1-\alpha)x_{t}$}, a linear combination of the current prediction and a recent solar current measurement ($x_{t}$) that is weighted by a parameter $\alpha$ which we set at $0.15$ as outlined in \cite{hsuISPLED2006}.
EWMA as implemented in \cite{hsuISPLED2006,kansalDAC2006} focuses on hourly predictions and utilizes historical data from the same hour on past days.  Our focus in this paper is primarily on daily solar current prediction, so we    adapt EWMA  to use  daily  historical data, noting that this algorithm was originally geared toward using hourly historical data.  Our adaptation uses the past set of daily historical data instead of the same day on past years due to the lack of sufficient historical data for the same day of the year. We believe our modified EWMA provides a fair comparison point for our models in this paper.

In comparing these models, for the winter data set, we see the results in Table \ref{table:environResults}, which shows our MLR model based only on environmental data improves over Persistence by $15\%$ and improves over EWMA by $16\%$.
For the summer data set, we similarly can compare the environmental only MLR model to Persistence and EWMA, seeing improvements of $19\%$ and $15\%$, respectively.
The residual mean indicates that our model has a tendency to underpredict in summer; on average this underprediction is 1.4mA with a 3.5mA confidence interval around this mean.
EWMA and Persistence overpredict in summer, as seen by the negative mean values, and both have a larger (compared to our model) confidence interval around their means at 4.2mA and 4.4mA respectively.
In winter, our model has a residual mean of 1.4mA, which is also an underprediction, and the confidence interval around this mean is 4.5mA.
EWMA and Persistence both also underpredict on average at a value less than our model, but with a larger confidence interval around that band that suggests more variability in those predictions.
Overall, these results suggest that our model has a better error bound on its results with a consistent underprediction that ensures the system continues to operate.
Overpredicting would allow the system to use more energy than it actually accumulates, thus decreasing the network lifetime; underpredicting ensures the system has energy remaining after the day ends and will lead to an increase in the network lifetime.

The different behaviors between winter and summer reflect the occurrence of more outliers in winter.
This winter had several days of abnormally high solar current in July along with a several periods of flat solar current in June and August.
Independent of whether these behaviors accurately reflect the weather or indicate startup issues with data collection, the models still capture the overall behavior and bound the error within reasonable ranges.

We also use the summer data set to explore the impacts of neighboring solar measurements, thus providing a spatial aspect to the data, as well as exploring a combination of neighbors' data and environmental data.
Utilizing neighbors' measurements could provide more richness to the solar data by indicating regional trends or help in situations where environmental sensors do not exist.
Our results show that environmental data alone provides the best predictions for the MLR model with improvements of $4\%$ over neighbors' only and $5\%$ over the combination.
Intuitively, the combination may be providing too much data as we then require the model to use both neighbors' data and spatial  data leading to an overfitting issue.
However, these percentages are quite small while these two variations on the MLR construction improve over Persistence by $16\%$ and $14\%$, respectively, and over EWMA by $12\%$ and $10\%$.
This indicates that any of the different structures of the MLR model will provide  improved predictions compared to existing methods.

\begin{table*}[h]
\centering
\tbl{Results of Different Models Predicting Daily Average Charge Current for Data Set 2008\label{table:environResults}}{
{\small
\begin{tabular}{|c|p{6.5cm}|c|p{1.6cm}|p{1.5cm}|p{1.4cm}|}
\hline
\multicolumn{2}{|c|}{\bf Model Type} & {\bf RMSE} & {\bf Maximum Absolute Error} & {\bf Mean Residual} & {\bf 95\% Confidence Interval} \\ \hline \hline
\bf{Winter} & & & \\ \hline
Our Model & 3 Solar, 1 Wind Direction, 1 Wind Speed, 1 Leaf Wetness, 1 Soil Moisture, Use Prediction Error, Recalibrate & 24.41 & 10.26 & 1.43 & 4.45 \\ \hline
Persistence & & 28.06 & 18.44 & 0.64 & 5.85 \\ \hline
EWMA & & 29.13 & 15.22 & 0.71 & 5.58 \\ \hline \hline
\bf{Summer} & & & & & \\ \hline
Our Model: Env. Only & 1 Solar, 1 Wind Speed, Use Derivative & 16.67& 3.40 & 1.74 & 3.54\\ \hline
Our Model: Spatial Only & 3 Solar, 10 Node21, 1 Node7, 9 Node9, Use Derivative & 17.33 & 4.75 & 2.78 & 3.29 \\ \hline
Our Model: Env. \& Spatial & 1 Solar, 10 Node21, 8 Node7, 7 Node9, 1 Wind Speed, Use Derivative & 17.63 & 4.29 & 3.10 & 3.33 \\ \hline
Persistence & & 20.59 & 6.33 & -0.60 & 4.39 \\ \hline
EWMA & & 19.68 & 5.23 & -0.57 & 4.19 \\ \hline
\end{tabular}
}
}
\end{table*}

\begin{figure}[htb]
\centering
\includegraphics[width=0.5\columnwidth]{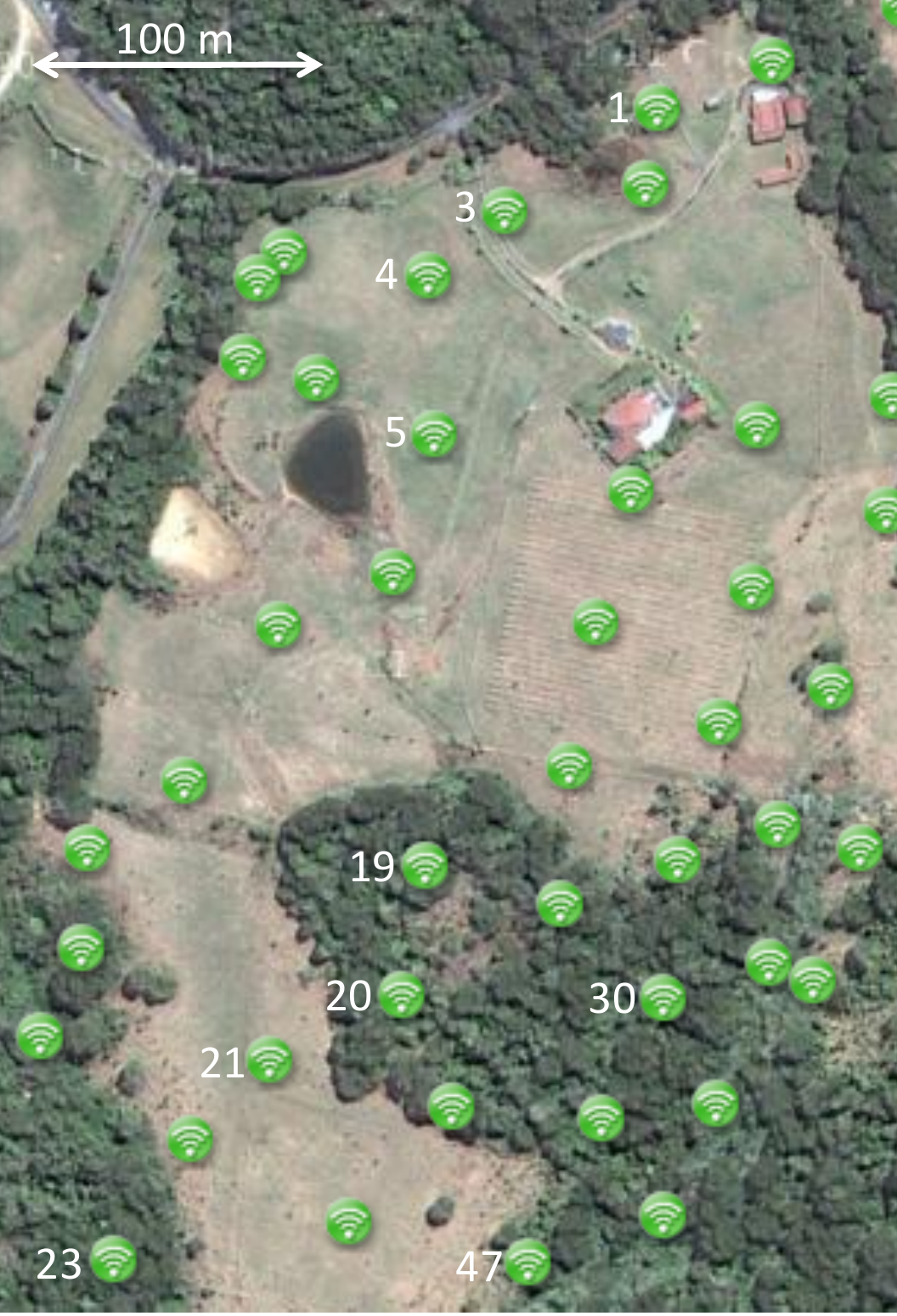}
\caption{Springbrook 2011 Deployment with Cluster Nodes Indicated by Node ID}
\label{fig:deploymentSite}
\end{figure}

\subsection{Spatial Analysis}
Given these results,  we now focus on spatial data as the most general case that does not require any additional sensors at the node for solar prediction. We explore the utility of neighbors' solar current measurements within the model to better understand how these nodes aid the prediction.
We examine the relationship between solar current predictions for nodes in different locations, which we classify as either {\it sunny}, {\it shady}, or {\it border} (a node at the edge of the canopy cover).
Given a node located in a sunny region, we want to know if it is better for that node to only use its own measurements, a collection of other sunny nodes, a collection of shady nodes, or some  mixed combination.
Here we utilize a more recent data set from 2011 with over 50 nodes and select representative nodes from our different classifications.
Our selection of clusters consists of representative clusters and not all possible clusters; in choosing nodes, we focused on ensuring disjoint clusters with clear classification labels.
This resulted in 10 different nodes used for our analysis that somewhat cover the region and limits the use of border nodes (which we could misclassify).
Figure~\ref{fig:deploymentSite} indicates the locations of these nodes in the region.

\begin{figure}[htb]
\centering
\includegraphics[width=0.8\columnwidth]{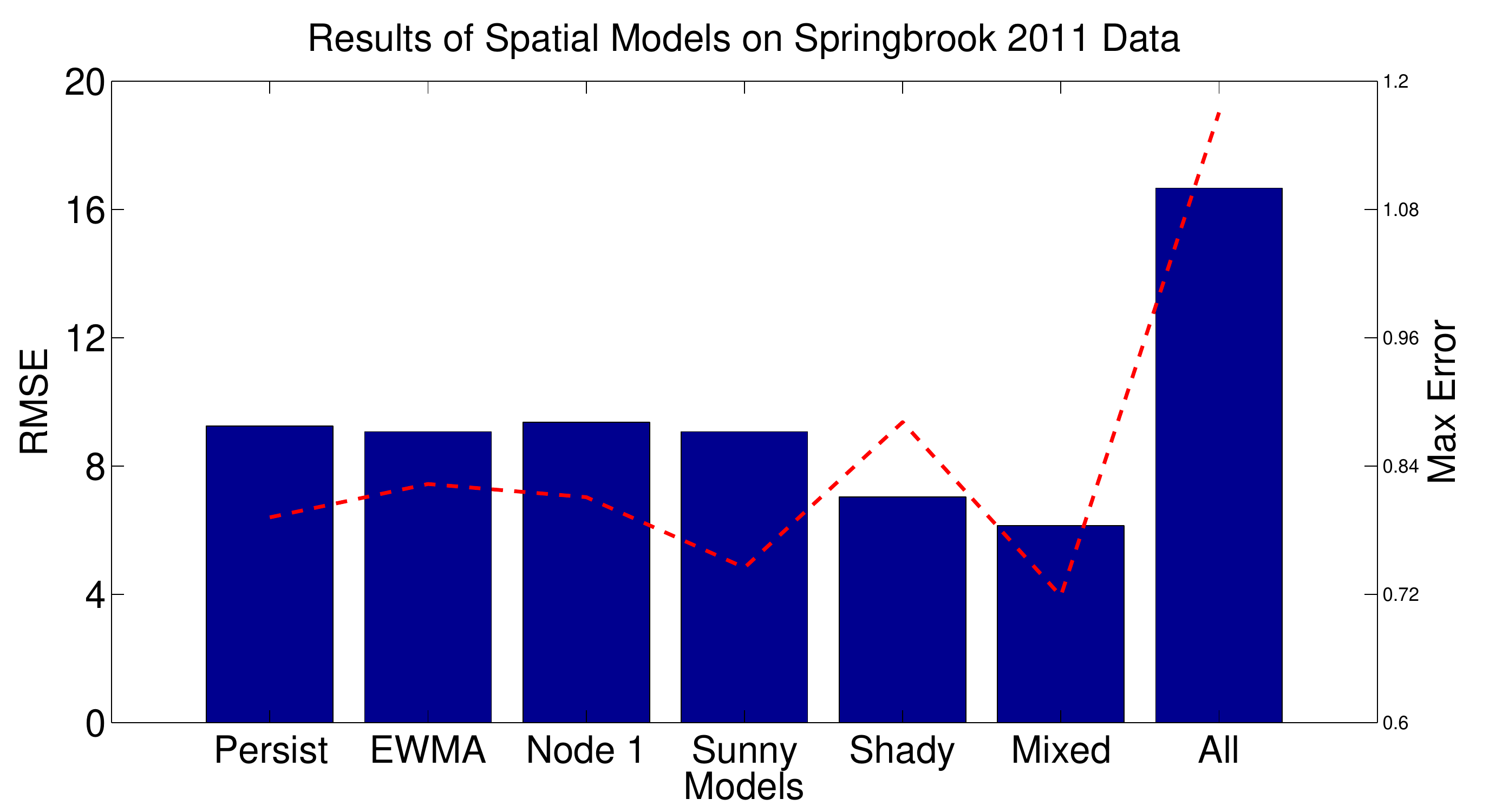}
\caption{Results of Spatial Analysis: Note that bars represent RMSE values and the dotted line represents Max Error}
\label{fig:spatialAnalysis}
\end{figure}

Using Node 1 as the primary node, we predict two days in advance using seven days for calibration as computed for the 2008 data.
Figure~\ref{fig:spatialAnalysis} shows the results for Node 1 predicting only using its own data and the MLR model, Persistence, EWMA, a collection of three neighbor clusters over various locations types, and using all nine neighbors from those clusters.
For our neighbor clusters, {\it sunny} includes nodes $[3, 4, 5]$, {\it shady} includes nodes $[19, 20, 30]$, and {\it mixed} includes nodes $[21, 23, 47]$.
 The results show that our MLR model can reduce the RMSE by up to one third over previous methods when a diverse group of nodes is involved in the prediction, while grouping shady nodes
 together to perform the prediction yields  a slightly smaller RMSE improvement and a higher maximum error. Self prediction and homogeneous sunny groups using MLR performed comparably to Persistence and EWMA in terms of RMSE, but sunny groups do reduce the maximum error.    Performing the prediction with all nine neighbors  has significantly higher  error. These results suggest that solar prediction  for a single node based on a small number of  neighboring nodes with spatial and solar measurement diversity is beneficial.
Our future work will explore automating the selection of which nodes to include in clusters.
A possible first order solution is to choose nodes by physical proximity in situations where the network topologies clearly cluster without issues of overlap.

\section{Testing: Implementation} \label{sec:solar-implement}
We implemented our algorithms on a physical sensor network platform to verify their functionality and feasibility in a real world scenario.
This section first describes the hardware platform and implementation of the algorithms on that platform, followed by the first field algorithm and its results, the second field algorithm and its results, and an analysis of the energy requirements based on data collected from the experiments.

\subsection{Fleck Platform} 
The implementation uses the Fleck\texttrademark3b platform for empirical validation.
The node is a low power wireless sensor network device designed by CSIRO specifically for outdoor applications such as environmental monitoring. 
The Fleck\texttrademark3b employs the ATmega1281 microprocessor running at 8MHz with 4K bytes EEPROM, 8K bytes SRAM, and 128K bytes program flash.
This low power microcontroller is combined with the Nordic NRF905 digital transceiver, which enables the Fleck to communicate at 50Kbps across 1km with a 915MHz quarter-wave whip antenna while consuming less than 100mW of power.
This platform can sense onboard temperature and power usage and is easily interfaced to numerous external sensors via the external sensor connector block and the daughterboard expansion bus.
On the software side, it runs Fleck OS (FOS)~\cite{corkeEWSN2008}, a cooperative threading operating system designed specifically for sensor networks.

For this installation, we used Fleck\texttrademark3b nodes powered by AA batteries that recharged with a 2.4 Watt solar panel~\cite{Corke_pieee10} as shown in Figure~\ref{fig:node}.
The nodes were specifically designed for validation of the power systems and communication systems so did not have environmental sensors.

In terms of the communication underlying our algorithms, we use Remote Procedure Calls (RPC) as described in~\cite{Corke_pieee10} for communication among the nodes.
Each node A broadcasts an RPC request to the next node B in the chain and waits to receive a reply.
Once B sends the reply, it assumes responsibility for the next step in the algorithm.
Node A receives the RPC reply to learn that responsibility has passed to node B.\newline

\subsection{Algorithm Implementation}
We implemented the algorithms of Section \ref{sec:solar-model} on the network with the algorithms running directly on the Fleck\texttrademark3b nodes.
The algorithms ran on top of the Fleck OS software and fit within the node's program memory.
On the MLR prediction side, the sensor nodes average the measured solar current over the day and then perform the MLR prediction computation as outlined in Algorithm \ref{alg:distPred}.

Algorithm \ref{alg:distCal}, the calibration of the model, requires a more complicated implementation.
To start, each node knows what columns of the matrix it maintains, granting the node with column $0$ master control of the operation.
We will denote each node as $x_i$ where $i$ defines the order each node has in storing the columns of the matrix (i.e. $x_0$ stores the first set, $x_1$ stores the second, etc.).
While for this test we fixed these values, the nodes could dynamically decide this placement as well.

Each node runs a state machine-like control loop defining where the node is in the algorithm.
The control loop begins in the waiting state with $x_0$ determining when to start based on a defined calibration window.
Node $x_0$ then transmits a command to load the data for the start of the algorithm causing all nodes to transition to the QR algorithm state, where each operates as outlined in Algorithm \ref{alg:qr}.
Upon completion of the QR algorithm, $x_0$ requests the $R$ and the $Q_i^Tb_i$ values from the other nodes and commences the SVD state.
The other nodes return to the waiting state at this point.
Node $x_0$ then completes the SVD state and pseudoinverse state, concluding with a transmission of the new coefficients for the MLR prediction algorithm to the other nodes.
All nodes save their portion of the coefficients and $x_0$ transitions to the waiting state, completing the algorithm.
Once the next calibration window occurs, due to any number of policies such as an increase in the prediction error or a scheduled cyclic calibration event, the algorithm and state machine begin again.

In addition to implementing these algorithms, we ensured fault detection, correction, and tolerance in this implementation.
The network will handle any of the following issues:

\begin{itemize}
\item All zeros in the first column data
\item Q columns received out of order
\item Q column not received
\item R column not received
\item Coefficients all zero
\end{itemize}

If the data is all zeros in the first column (indicating no solar current measurements exist yet or potential errors with the solar charging system), the algorithm will not commence and will retry later.
For errors where columns are not received or arrive out of order, the algorithm will fail gracefully and retry later.
If at the end of the algorithm, the coefficients received are all zeros, the nodes will not load these values, but maintain the old values and wait for a retry of the algorithm.
Currently there is no data replication to ensure successful completion of the algorithm upon complete failure of a node; we leave this for future work, but believe there are many simple policies to fix this. 

The final program on the Fleck occupies 36,842 bytes in program flash, which is well within the 128K flash memory of the platform. Our experiments use node group sizes of up to 5 nodes, which  means that $m=5$, and corresponds to a RAM footprint of $5*5*32$ bytes $=800$ bytes (as each solar value is 32 bytes).  This RAM requirement does not present an issue for the 8K RAM on the Fleck.

\subsection{Experiment 1}
Our first field experiment addresses proof-of-concept questions and proves the overall operation of the system.
For the experiment we used three nodes placed on the CSIRO ICT campus as shown in Figure~\ref{fig:node}.

\begin{figure*}[htb]
\centering
  \subfigure{\includegraphics[width=0.75\columnwidth]{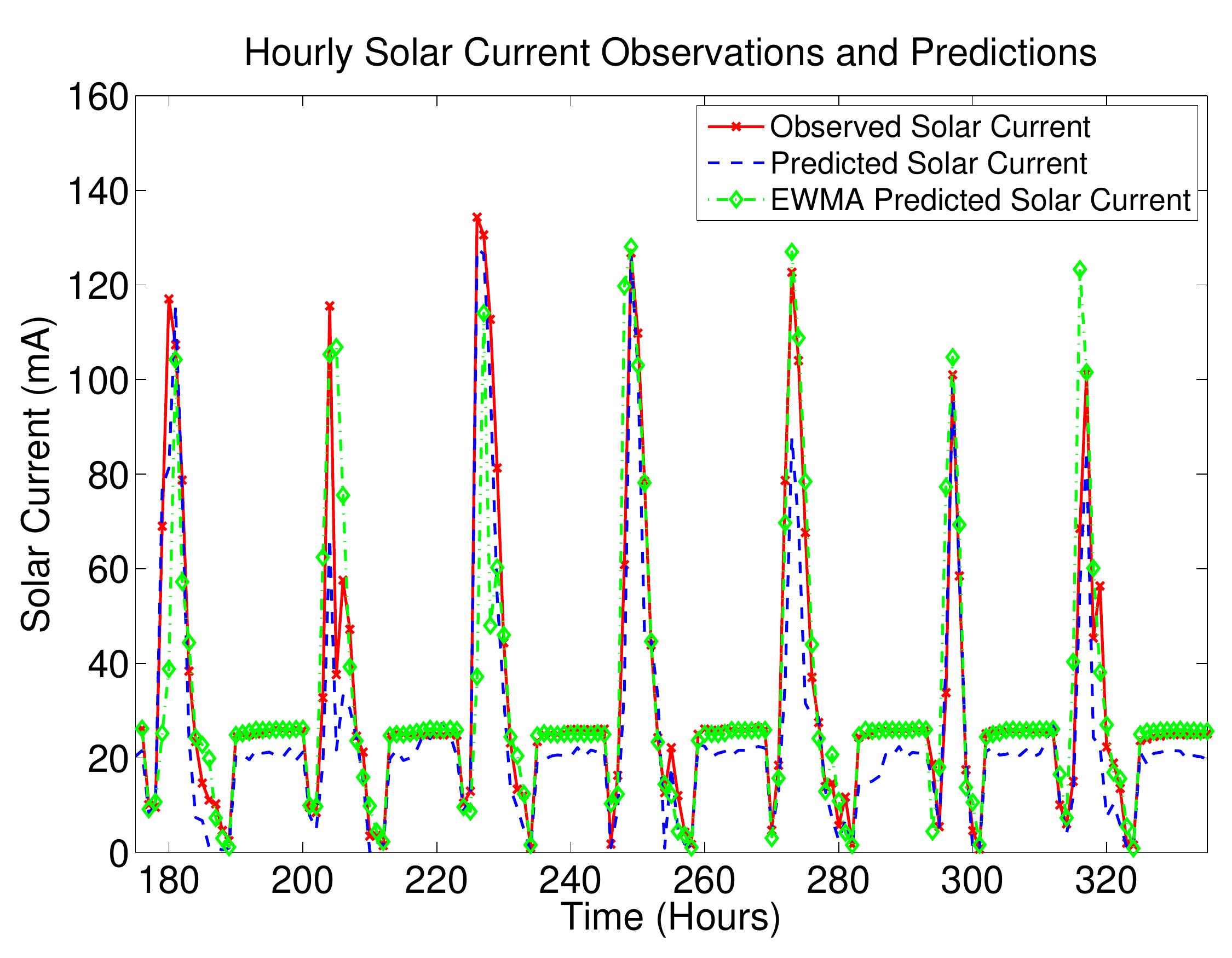}\label{fig:hourlyData_Results1Week}}
  \subfigure{\includegraphics[width=0.76\columnwidth]{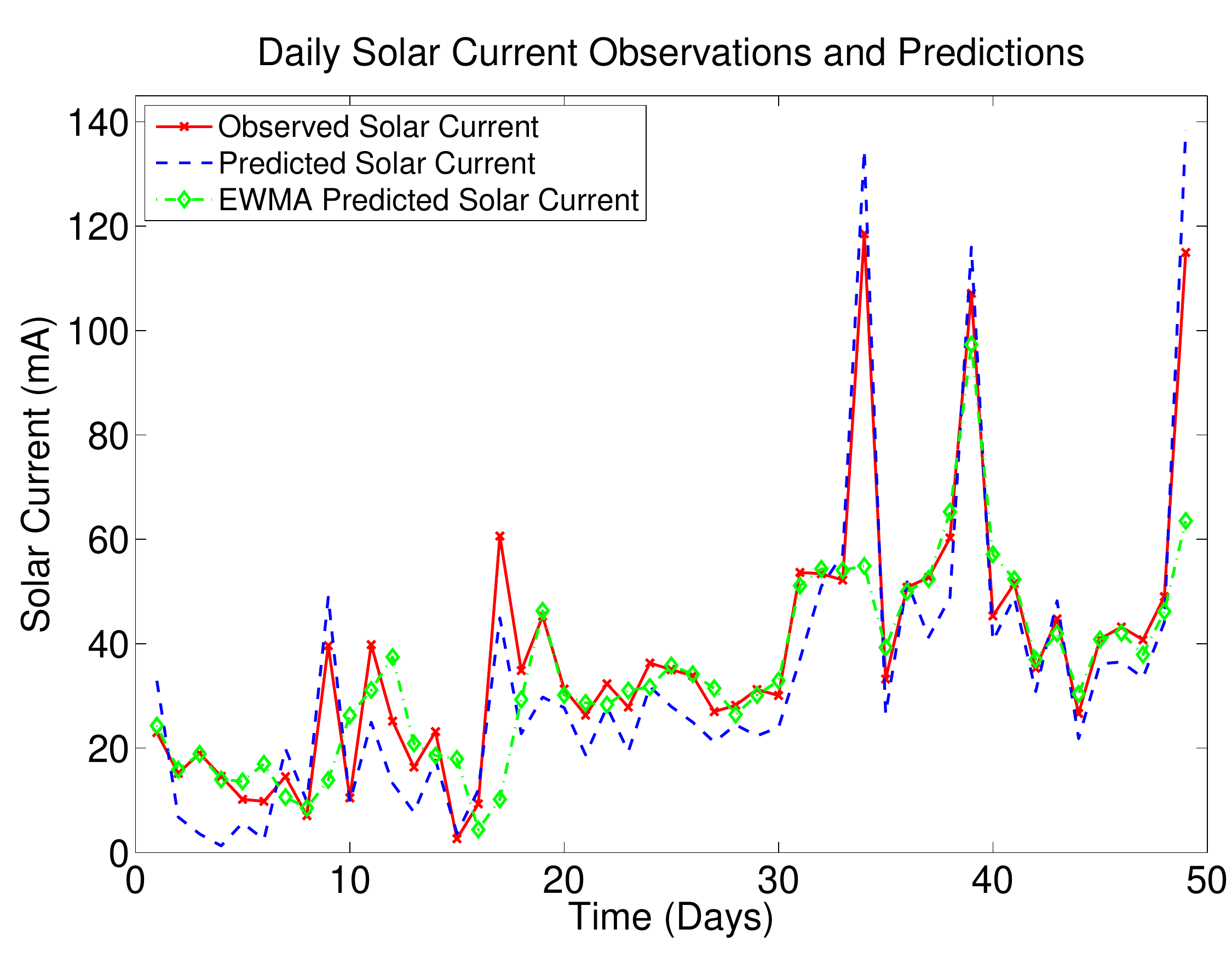}\label{fig:dailyDataResults}}
\caption{(a) One Week of Observed and Predicted Data from Test 1 and (b) Daily Average Solar Current Observed and Predicted from Test 1}
\label{fig:dataResults}
\end{figure*}

We ran the system for over seven weeks.
Initially, the system measured every 15 minutes, predicted every 15 minutes, and calibrated every 90 minutes.
The predictions predicted the solar current two measurement intervals in the future, or 30 minutes.
Figure \ref{fig:hourlyData_Results1Week} shows one week of the data from this time period, aligned and averaged to the hour.
The prediction performs reasonably well compared to the observations.
An interesting phenomenon occurs in the graph with two sharp down spikes between the daytime period and the nighttime period.
At night, the system actually is measuring the nighttime lights that illuminate the campus (an unforeseen effect).
The two spikes then reflect when those lights turn on and off.
Our predictions capture this unusual behavior as well.

After approximately three weeks, we changed the system parameters to test the calibration algorithm more rigorously.
We shortened the measurement window to two minutes and the calibration window to six minutes.
This also shortened the prediction window to four minutes (as the system still defined it as two measurement intervals).

Over a two week period, the system attempted 2773 calibrations.
2679 of these were successes and 94 failed, resulting in a $96.6\%$ success rate.
The failures all occurred early on as a result of one node having a low initial battery voltage.
This node shut down until the next morning when it received sufficient solar energy to resume operation.
However, the system correctly identified the failures when the node did not contribute to the algorithm with the remaining active nodes timing out and returning to the ``wait'' state of the control flow.
They continued to attempt calibration and eventually succeeded, all without user intervention.
These instances of low power states, while not ideal for our operation, also argue the need for smart energy management and the usefulness of our system.

To connect the test to our daily average predictions from simulation, we analyzed the overall data set to see how well it predicted the daily average solar current.
We averaged both the observations and predictions; Figure \ref{fig:dataResults} compares the observations, the MLR predictions from the system, and an offline prediction by the EWMA model.
Our predictions match the  observations well and better capture the peaks compared to the EWMA model.

\begin{table}[htb]
\centering
\tbl{Results of Models Predicting Average Hourly and Daily Charge Current for Data from CSIRO Test\label{table:csiroRMSEresults}}{
{\small
\begin{tabular}{|p{0.2\columnwidth}|c|c|c|c|} \hline
 & \multicolumn{2}{|c|}{\bf Hourly} & \multicolumn{2}{|c|}{\bf Daily} \\ \hline
{\bf Model Type} & {\bf RMSE} & {\bf Max Abs Error} & {\bf RMSE} & {\bf Max Abs Error} \\ \hline
Our Model & 2.04 & 0.98 & 0.91 & 0.91\\ \hline
Persistence & 3.61 & 22.33 & 2.52 & 7.70\\ \hline
EWMA & 1.95 & 4.93 & 1.51 & 5.75 \\ \hline
\end{tabular}
}
}
\end{table}

We also analyze the RMSE error and maximum absolute error for the hourly and daily averaged data sets.
Table \ref{table:csiroRMSEresults} lists these values, which verify the graphical analysis.
Regarding EWMA in this table, hourly uses the model as described in~\cite{hsuISPLED2006,kansalDAC2006} while daily uses the modified version described in Section~\ref{sec:simulation}.

For hourly predictions, we see the benefit of the specialized historical data in the EWMA model as it provides the lowest RMSE error with a $4.5\%$ improvement over our model.
However, EWMA exhibits a $403\%$ increase in maximum absolute error over our model. 
Our model greatly improves over Persistence by $44\%$ for the RMSE and $95.6\%$ for maximum error.
For daily predictions, our model has the lowest RMSE and maximum absolute error compared to both Persistence and EWMA.
It reduces the RMSE by $64\%$ and $40\%$ over Persistence and EWMA, respectively, and the maximum absolute error by $88.1\%$ and $84.1\%$.

These results indicate that our model works well over time, as seen by the RMSE numbers, and instantaneously, as seen by the maximum error numbers.
While the intent of our model is for larger daily time windows, it also achieves reasonable performance at the hourly time frame.
These numbers could improve further by incorporating more diverse neighbor data as we explored in Section~\ref{sec:simulation}.
Additionally, our model underpredicts as seen in the graphs, providing a conservative estimate of future solar energy.
This leads to the model saving power.
An overprediction could mislead the nodes into thinking more energy exists than really does, thus implementing a more power hungry mode of operation and leading to node failure.
By underpredicting, the system utilizes less power than necessary, ensuring reserves exist for overly cloudy and rainy weather.

Overall, this test demonstrated a functional and correct implementation of the algorithm on the platform, connecting the theory to the real operation of the algorithms.\newline

\subsection{Experiment 2}
To scale the experiment up and examine spatial coverage issues, we performed a second field experiment of 10 nodes operating over four months.
The nodes were divided into two five-node groups and placed at different campus locations with different solar exposure.
Each group had its own master node and consisted of nodes within one-hop range.
The nodes also transmitted their data out of the network to the base nodes and gateway located on the campus.
These nodes did not participate in the computation, but received the data and recorded it to a database with web access.

Nodes measured solar current every five minutes and averaged the data over 15 minute intervals.
The master predicted the future solar current 30 minutes ahead with calibration occurring every 1.5 hours.
This provided enough events to ensure functionality without overloading the communication system.
After setting the experiment up, we monitored it sporadically, verifying that our algorithm functioned properly in a larger network size.

During the experiment, the campus experienced quite a lot of stormy weather including one storm classified as a natural disaster, affecting the overall operation of the system at different points: power failures of the gateway node, central database corruption, and limited solar recharging, to name a few.
From these experiences, we determined that we need better energy management.
We had not connected our prediction to the actual energy management of the node; if we had, we might have reduced measurement windows and other behaviors to conserve energy and ensure continual operation during the storms.

However, when the nodes had sufficient power, they did continue to run the algorithms.
Based on the overall counts of 766890 calibration attempts and 82258 errors, we saw a $89.3\%$ success rate in computing the calibration algorithm over the entire time period.
Overall, this demonstrated that we can scale the algorithms to larger node sizes and that more work is needed in linking outputs of the prediction to an energy management strategy. \newline

\begin{table*}
\centering
\tbl{Energy Requirements for Models\label{table:csiroEnergyResults}}{
{\small
\begin{tabular}{|c|p{3.0cm}|p{2.5cm}|p{3.0cm}|p{2.9cm}|}
\hline
{\bf Model Type} & {\bf Computation Ops} & {\bf Computation Energy (mJ) } & {\bf Communication Msgs} & {\bf Communication Energy (mJ)} \\ \hline
Our Model: MLR Prediction & 5 & $2.38\times10^{-4}$& 2 & 1\\ \hline
Persistence: Prediction & 0 & 0 & 0 & 0 \\ \hline
EWMA: Prediction & 3 & $1.43\times10^{-4}$ & 0 & 0 \\ \hline
Our Model: Calibration & 8400 & 0.4 & 6 & 3.1\\ \hline
\end{tabular}
}
}
\end{table*}

\subsection{Energy}
Despite the improved predictions, we must ask whether it makes sense from a energy standpoint to perform a more complex model.
Table \ref{table:csiroEnergyResults} outlines the relevant numbers.

First, from the prediction side, computationally, Persistence performs the best as it requires no computation while our model and EWMA differ by two operations, leading to a $9.5\times10^{-5}$mJ increase in energy in order to compute our model.
Distributing the MLR prediction for our model requires two 32-byte messages, which requires 1mJ of energy, while the other methods have no communication requirement and, thus, require no energy.

On the calibration side, only our model requires calibration.
This calibration, using the parameters of our field test, requires roughly 8400 operations (only 6.7ms of processing time).
On our 8MHz Fleck platform operating at 3.3V and requiring 18mA active current, this results in 0.4mJ of energy.
Communicating the messages required for calibration consists of six messages at a maximum transmit power of 100mW and data rate of 50Kbps, requires 3.1mJ of energy.
To put these numbers in perspective, during our field test, the system gathered 37.5mA daily (on average).
This resulted in $1.07\times10^{4}$J of energy daily (296.7Wh).
As the trend for our model is underpredicting, we would easily recover the energy costs of computing our model through the energy savings incurred by using the prediction to manage power.
For instance, if the sensor node sets the sensor sampling rate based on the predicted energy that it will have available, using our model allows the node to set the sampling rate more accurately for its energy budget. 
In return for this energy expenditure, we see a $39.7\%$ improvement in our solar current prediction over EWMA and a $63.9\%$ improvement over Persistence.
Table \ref{table:resultSummary} outlines the key results.
The significant improvement in prediction accuracy for a small energy overhead confirms the benefits of our distributed regression model for solar prediction and more informed energy management on sensor nodes. \newline

\begin{table}
\centering
\tbl{Summary of Key Results \label{table:resultSummary}}{
{\small
\begin{tabular}{|c|c|}
\hline
Daily Energy Harvested & $1.07\times10^{4}$J \\ \hline
Extra Energy Required & 4.5mJ \\ \hline
Prediction Improvement & 40-60$\%$ \\ \hline
\end{tabular}
}
}
\end{table}

\section{Discussion and Future Work}\label{sec:solar-discuss}
Based on these results, we elaborate on several aspects of the system.

First, we saw single node failures during our testing, but avoided multiple node failures.
What would happen if a multiple node failure occurred?
Should any nodes other than the master node fail, the network would continue to attempt calibration, miss the needed communication, and cancel the rest of the algorithm.
Should the master node fail, the calibration would not occur.
Future work entails adding data redundancy to the algorithms such that the failure of a node still allows the calibration algorithms to continue.

Another thing to note regarding these results is that the failure of the calibration algorithms does not mean the failure of the system.
In cases where the calibration fails to complete, MLR prediction does continue with the old coefficients.
Energy management can still proceed with predictions of future solar current; however the coefficients may be less accurate.
Slightly less accurate predictions will not cause our network to fail.
Any time series prediction will oscillate around the real values, but the aggregate behavior should match the real solar current behavior; this is what we see in our work.
As long as the aggregate matches and predictions do not oscillate too far from the real behavior (oscillations which we do not see in our work), any energy management system will be able to maintain operation.
Some days will require using more energy from the power storage system than expected, but this will be corrected on those days where extra power is stored.
Overall, the predicted solar current will lead to a more comprehensive energy management system, which should ensure the continual operation of the network.

Additionally, we can explore the trade-offs regarding the time windows for sensing, prediction, and calibration.
All the timing is sensitive to the phenomena and data being predicted, but we can make some generalizations.
Small time windows for sensing and prediction occasionally result in models too sensitive to perturbations within the data; the system then requires more frequent calibration to adjust and some amount of smoothing.
For the case of solar current at a medium time scale (approximately of the order hours), the cyclic nature of the data with high values during the day but near zero values at night can confuse the model, ensuring it does not accurately predict either.
In our results, Figure \ref{fig:hourlyData_Results1Week} sees the model not quite matching the nighttime values and, on occasion, not quite reaching the heights of the daytime values.
Accurate models at these medium time scales might require operating only with one portion of the cycle, but the movement of the transition points between night and day and vice-versa can still cause difficulties.
Larger time scales, such as the daily values we use, smooth over the cycles and perturbations to ensure a data set more amenable to time series predictions.
However, this does require more time to gather data for calibrating and recalibrating based on seasonal trends in the data.
Our choice of daily time windows ensures a reasonable prediction while minimizing the amount of computation (and thus power usage) needed by the model, allowing more time for monitoring the environment and the operational goals of the sensor network.

We also want to consider the definition of the matrix and determination of appropriate policies.
We used the existing simulation data to define reasonable model structures and chose fixed windows for calibration and MLR prediction.
Ideally, the system would dynamically determine this, either in a central or distributed approach, allowing for a more adaptable model.
This would also allow for a variety of prediction windows and dynamic growth of the calibration window as more data arrives.
With dynamically defining these parameters, we need to also decide the optimal strategy for predicting, whether it makes more sense to utilize the same prediction for a node cluster or have each node predicts its own current.
This relates to node density and the variability of the environment.
A more heterogeneous setting may lead to better individual predictions while a homogeneous setting to cluster predictions; also, a large network most likely will need a combination of approaches.
To achieve this, we first must include manual methods for defining these policies and then consider automated methods.
We leave this for future work, recognizing the importance of answering these questions to enable a portable prediction model.

Finally,  an interesting direction for future work  is to further explore the relationship between prediction accuracy on one hand and the spatial and measurement diversity on the other.
Our results from Figure \ref{fig:spatialAnalysis} indicate some interesting opportunities to improve estimation quality through the grouping of  small groups of nodes with diverse solar measurement profiles. Further study is required to establish persistent trends regarding the impact of the number of nodes in a prediction group and the degree of diversity in their solar measurements on  the solar prediction accuracy.

\section{Conclusions} \label{sec:solar-conc}
Optimizing energy usage on sensor nodes is a key issue for sensor networks.
In this paper, we describe a model and distributed algorithms for predicting future daily average solar current.
We develop a distributed pseudoinverse algorithm usable in a wide range of applications.
We verify the functionality of these algorithms through simulation and a month-long field experiment on our platform.

Predicting solar current enables better power management in sensor networks.
For example, a power management system could use the solar current prediction to determine that insufficient energy will be harvested over the next two days to support the current operations.
A power management planner could use this information to plan the system operation with fewer communication rounds, less data, or perhaps compressed data, trading off power and communication for computation, data resolution, and solution accuracy.
Overall, better power management means longer operation of the network, providing more monitoring of the environment, more data, and a more useful sensor network.
Our work equips sensor networks to provide better energy management and does so in a very usable, general form. 
Anyone can use our algorithms to predict future energy on any sensor network with solar recharging.
The model utilizes any combination of climactic and spatial variables available on the platform; the only requirement is the system measures its solar current.

Moving forward we will explore dynamic model definition, allowing better adaption of the models to changing networks and new installations.
We will also explore the scalability of the model to larger sized networks and further approaches to improved energy prediction to ensure longer lived networks.

\section{Acknowledgments}
We would like to acknowledge the following groups for their support: CSIRO, the Australian Academy of Science, and the National Science Foundation EAPSI grant number 6854573.
This work is also supported in part by ONR grant number N00014-09-1-1051 and CSIRO's Sensors and Sensor Networks Transformational Capability Platform.
We would also like to thank Ben Mackey for his help with running the experiments.


\received{August 2011}{January 2014}{March 2014}

\end{document}